# Nonlinear vs. bolometric radiation response and phonon thermal conductance in graphene-superconductor junctions

Heli Vora, Bent Nielsen and Xu Du

Department of Physics and Astronomy, Stony Brook University, Stony Brook, NY

## Abstract

Graphene is a promising candidate for building fast and ultra-sensitive bolometric detectors due to its weak electron-phonon coupling and low heat capacity. In order to realize a practical graphene-based bolometer, several important issues, including the nature of radiation response, coupling efficiency to the radiation and the thermal conductance need to be carefully studied. Addressing these issues, we present graphene-superconductor junctions as a viable option to achieve efficient and sensitive bolometers, with the superconductor contacts serving as hot electron barriers. For a graphene-superconductor device with highly transparent interfaces, the resistance readout in the presence of radio frequency radiation is dominated by non-linear response. On the other hand, a graphene-superconductor tunnel device shows dominantly bolometric response to radiation. For graphene devices fabricated on $SiO_2$ substrates, we confirm recent theoretical predictions of $T^2$ temperature dependence of phonon thermal conductance in the presence of disorder in the graphene channel at low temperatures.

## Introduction

Bolometers are sensitive electromagnetic radiation detectors, which measure the incident radiation power through radiation-induced heating. In recent years, graphene has been studied as a promising nanomaterial for bolometer applications [1-5], with potential advantages including its extremely small heat capacity [2] and weak electron-phonon coupling at low temperatures [6-9]. The small electronic heat capacity, resulting from the ultra-small volume of the material combined with the low electron density of states, can be estimated to be $C_e \sim 10^{-21}$ J/K for a 1 μm$^2$ graphene flake at T∼4 K and a typical carrier density of $10^{12}$ cm$^{-2}$ [4]. The weak phonon coupling makes it possible to achieve low thermal conductance (G) and hence high intrinsic sensitivity, as characterized by the thermal fluctuation limited noise equivalent power $NEP_{th} = \sqrt{4k_B T^2 G}$. While details of the device design requires careful balancing of various parameters [3, 10, 11], it is generally acknowledged that by combining small heat capacity and low thermal conductance, graphene offers promise for building bolometers which simultaneously achieve high sensitivity and fast response with a small time constant.

A practical graphene-based bolometer built to utilize the above advantages should satisfy several criteria. To achieve a low thermal conductance, cooling of the hot electrons from diffusion (through the contacts) must be suppressed to be lower than the phonon cooling. For a highly doped, disorder-free graphene flake, the thermal conductance due to electron-phonon scattering scales with the electron temperature as $G_{ph} = 4\Sigma A T_e^3$ where Σ is the electron-phonon coupling and *A* denotes flake area. [8, 9] The thermal conductance from diffusion can be estimated based on the Wiedemann-Franz law, $G_{WF} = \alpha L T / R_s$ where $R_s$ is the total contact resistance ($R_s = 2R_c$, $R_c$ being the DC resistance of each contact), α is a geometrical factor and L is the Lorentz number equal to $2.44 \times 10^{-8}$ WΩK$^{-2}$. For a graphene-bolometer with a size of 50 μm$^2$, for example, one can estimate that a contact resistance of ~100 kΩ at 1 K and ~7 kΩ at 4 K is needed to satisfy $G_{WF} \ll G_{ph}$, if non-superconducting contacts are used. (Here, Σ is taken as 50 mW/K$^4$m$^2$ [6-9] for a Fermi energy of 100 meV and $\alpha = 4$ for the typical two-terminal device geometry).

Coupling efficiency needs to be considered as well since typical graphene-bolometer devices are much smaller than the radiation wavelength (e.g., RF or THz) and an antenna is required to couple them to the radiation. Therefore impedance matching with the antenna is important for high detection efficiency. To achieve efficient coupling, a low device impedance of 50 Ω is needed in the case of an amplifier. A 100 Ω device resistance is desired for coupling through planar antenna for high frequency signals.

Another important issue is how to detect the electron temperature change, as the resistivity of graphene itself depends only weakly on temperature [12]. A few attempts have been made towards solving this problem, including using semiconducting bilayer graphene [5], Johnson noise thermometry [1, 2] and graphene-superconductor junctions [4, 13].

Graphene-superconductor junctions can be designed to provide a promising solution to all of the technical challenges listed above. Although a high energy radiation pulse is absorbed by a few electrons, they quickly thermalize below the superconducting gap via electron-electron interactions, much faster than energy is given off to the lattice [5, 14]. The superconducting gap in the leads effectively confines these hot electrons, preventing hot charge carrier diffusion, while the contact impedance can, in principle, be made low enough for high coupling efficiency. In addition, the junction resistance has a strong temperature dependence, which can be used for a resistive readout apart from temperature detection through Johnson noise thermometry.

Two types of graphene-superconductor devices have been studied so far for bolometric response. One is a graphene-based Josephson junction, where Andreev reflection provides good thermal confinement and the supercurrent can be used as an electron temperature thermometer. While a large supercurrent has the disadvantage of over-heating when the device switches to normal state, it is in principle possible to design a graphene-superconductor junction with highly transparent interface yet with the supercurrent suppressed. As demonstrated later (see Fig 2(a)), the electron temperature could be detected through the DC resistance of the device, which is affected by the temperature-dependent penetration depth of the order parameter [15] inside graphene. Another type of device is a graphene-superconductor tunnel junction bolometer [4], where hot electron confinement is achieved by suppression of quasiparticle tunneling, and the tunneling resistance is used as an electron temperature thermometer.

In this article, we study the general scheme of radiation response of graphene-superconductor junctions, using superconducting contacts of Al and NbN. Two types of devices listed in Table 1, one with highly transparent graphene-superconductor interfaces (D1) and one with a large tunneling barrier at the graphene-superconductor interfaces (D2), are discussed. We find that with resistance readout, in graphene-superconductor junctions with transparent interfaces, microwave response is dominantly determined by the highly non-linear I-V characteristics. In graphene-superconductor tunnel junctions, on the other hand, radiation response is mainly bolometric. The tunneling devices show dynamic resistance, which decreases with increasing electron temperature. Due to a large tunnel junction capacitance these devices have low microwave impedance, allowing impedance matching and high microwave efficiency. As a result, graphene-superconductor junctions are promising for practical and highly sensitive bolometry.

Using graphene-superconductor tunnel junctions, we were able to study phonon cooling of the hot electrons in graphene. We show that, in agreement with the theoretical predictions [16] and the recent experimental study of electron-phonon cooling using Johnson noise thermometry [17, 18], the thermal conductance due to electron-phonon coupling follows a temperature law $G_{ph} \propto T^2$ in disordered graphene, quantitatively different from the $G_{ph} \propto T^3$ law for the clean limit.

**Non-Linear response in transparent graphene junctions**

Graphene-superconductor junctions with highly transparent interfaces offer unique advantages as radiation detectors. At the superconductor-graphene interface, hot electrons diffusing from graphene into the superconductor are converted isothermally into holes by the Andreev reflection process, as long as their energy is within the superconducting gap [19]. This provides thermal isolation of the hot electrons, reducing the diffusion contribution to the thermal conductance and therefore a high sensitivity can be achieved. Recently thermal confinement at highly transparent graphene-superconductor interfaces has been demonstrated with overdamped graphene-Pb weak links, by measuring the Joule heating-induced hysteresis in the I-V characteristics [13]. At sub-Kelvin temperatures, Pb electrodes isolate the graphene channel, so the dominant cooling mechanism is electron-phonon scattering. The thermal conductance observed is much less than what is estimated from the Wiedemann-Franz law. However, in these Josephson junctions with supercurrent, to use the junction resistance as readout, the bias current must be larger than the supercurrent. Once the Josephson junction switches to a resistive state, at low operating temperatures, the bias current may induce large unwanted heating and saturate the devices.

In our work, we focus on the possibility of using transparent graphene-superconductor weak links for bolometric detection of small microwave signals by directly using the DC resistance as the readout. In this device scheme, we choose to study the regime where supercurrent is suppressed by making the graphene channel relatively long and by tuning the chemical potential close to the charge neutrality point (CNP). Even in the absence of supercurrent, the advantages of low contact resistance (convenient for impedance matching and high coupling efficiency) and thermal confinement from the superconducting contacts are still present.

The devices were fabricated by mechanically exfoliating graphene on a doped Si/$SiO_2$ (285 nm) substrate. Contacts are defined using standard electron-beam lithography. Results discussed here are from junctions made with 2 nm Pd and 30 nm Al contacts, but we note that similar results were obtained with Ti/Nb and Ti/NbN junctions measured at higher temperatures. The results shown below are for a device with approximately 6 μm wide Pd/Al contacts separated by ~0.7 μm long graphene channel. Measurements were carried out by supplying a sweeping DC bias along with a small AC modulation of 50 nA at 13 Hz, both applied with a Keithley 6221 current source. The differential resistance (dV/dI) was measured using a lock-in amplifier, while the dc voltage was measured using Keithley 2182 nanovoltmeter. Measurements were done in a dilution refrigerator. The RF power was delivered to the device through a co-axial cable at approximately 1.29 GHz, where maximum device response was recorded. The co-axial cable was located about 10 cm away from the device and no antenna was used. This allowed a very small fraction of the applied power to be coupled to our device. Details of the measurement setup along with the method of applying RF power in this setup are described in our previous work on graphene-Al tunnel junction bolometers [4].

The high quality of the graphene-Al interface was confirmed by fabrication and characterization of a junction with leads separated by ~0.3 μm. At temperatures $T < T_c$ ($T_c \sim 1$ K being the superconducting transition temperature of Al), a well-defined supercurrent was observed. Together with the pronounced features of multiple Andreev reflections, these indicate high transparency of the graphene-Al interfaces. With exactly the same procedure, we fabricated and measured a device with slightly larger lead separation (~0.7 μm), where within a wide range of temperatures and gate voltages in the vicinity of the CNP, the supercurrent is absent. Fig 2(a) shows dV/dI vs. bias voltage curves at different bath temperatures, for the longer channel device away from the CNP. Even with no evident supercurrent we still observe a significant reduction in zero-bias resistance along with multiple Andreev reflections, again indicating high transparency of our contacts. From Fig 2(a), we see that the differential conductance changes with the temperature in the range $|V_b| < 2\Delta$ and is independent of the temperature outside the superconducting gap ($\Delta$). As shown in Fig 2(a) inset, by applying $V_g = V_{CNP} = -15$ V and by raising the bath temperature to 320 mK we effectively suppress the supercurrent. Here, contact resistance of only few tens of Ohms allows the device resistance to be tuned around 50 $\Omega$ by changing channel doping or temperature.

Characterization of the RF response in these devices is performed in a similar manner as reported in our previous work on bolometric response in graphene/TiOx/Al tunnel junctions[4]: by correlating the dV/dI vs. bias voltage curves in the presence of RF radiation vs. in elevated bath temperatures. The response to applied radiation in dV/dI vs. bias curves is shown in Fig. 2(b) bottom panel. At higher applied radiation power, we observe that the dV/dI vs. bias voltage curve develops a rather complicated behavior with oscillation-like features near zero-bias. Such features are not observed by heating the device through bath, where a temperature increase simply causes the curve to be more "smeared".

The origin of these features can be explained by the strong non-linearity in device response. The applied RF signal averages over a portion of the non-linear I-V curve to generate a voltage. Due to the nature of low-frequency lock-in amplifier measurements, this voltage mixes with the lock-in amplifier reference frequency. This kind of mixing produces low-frequency harmonics of the device response, which appear in the 13-Hz lock-in measurement. If this effect is greater or comparable to the device bolometric response, our detection scheme of overlapping curves at different bath temperatures with curves at different radiation fails. To test our hypothesis, we simulate the lock-in amplifier response to the device I-V curve taken at 320 mK, at $V_g = -15$ V, with no radiation. Lock-in amplifier response for our device I-V curve is given by,

$$V_{LIA} = \frac{2}{\tau}\int_0^\tau V_{device}\,(I_{applied})\sin(\omega_{lf}t)dt \qquad (1)$$

where, $\omega_{lf} = 2\pi \times 13$ s$^{-1}$ is the frequency of lock-in amplifier operation, $\tau$ is the lock-in integration time constant and $I_{applied} = I_0\sin(\omega_{lf}t) + I_{bias}$ in absence of a RF signal. (Here, $I_0 = 50\,nA$). We calculate the voltage developed in the device based on the input I-V curve information through interpolation at a range of applied currents $I_{applied}$, which is a superposition of a small low-frequency ac current $I_0$ on a DC bias current $I_{bias}$. By adding a high frequency current, $I_{RF}$, as an external input, we mimic the applied radiation. Then, the device response is calculated for $I_{applied} = I_0\sin(\omega_{lf}t) + I_{bias} + I_{RF}\sin(\omega_{RF}t)$. We can reproduce these non-linear features and see that the simulated curves overlap with the ones measured to a high degree inside the superconducting gap (Fig.2 (b)). The slight mismatch at higher biases and higher radiation power could be thought of due to the small bolometric device response that is not included in the simulation. Even though the onset of non-linearity induced features is at a high applied RF signal, if we take into account that only a fraction of this signal reaches the device due to the lack of antenna and conducting substrate losses, we can say that non-linear detection dominates the device radiation response when the device voltage is used as readout. Further exploration of graphene-superconductor junctions with highly transparent interfaces may instead focus on excitation-free readout techniques such as Johnson noise thermometry.

**Graphene-NbN tunnel junction bolometers**

In the superconductor tunnel junction (S-I-N-I-S) bolometer scheme, diffusion of hot carriers from the normal absorber into the superconducting leads through quasiparticle tunneling is largely suppressed and therefore a low thermal conductance can be achieved [4]. At the same time the strong temperature dependence of the tunneling resistance can be used as an electron temperature thermometer. Previously, we reported a bolometric response in graphene-superconductor tunnel junctions with TiOx tunnel barrier and Al contacts [4]. In these junctions, the RF response was identified to be bolometric by comparing the device resistance change with the microwave radiation absorbed with the resistance change due to changing of the bath temperature. Using the temperature dependence of the zero-bias differential resistance as an electron thermometer and by calibrating the absorbed radiation power in the device, we were able to obtain the bolometric parameters in these devices.

In this section, we discuss the general feasibility of the graphene-superconductor tunnel junction bolometer scheme and its unique advantages. In particular we try to understand the devices' RF response with higher $T_c$ superconductor contacts such as Nb and NbN than studied previously (Al) [4] and its coupling efficiency. We demonstrate here that: 1. these devices clearly show bolometric response under RF radiation, with higher sensitivity and phonon dominated cooling power compared to the devices demonstrated previously, with hot charge carrier diffusion dominated cooling power (with Al contacts); 2. their high frequency impedance matches well with the standard RF devices/antenna, which allows high coupling efficiency; 3. disorder in graphene plays an important role in determining the performance of these devices.

We deposit graphene on a high resistivity Si substrate (15-20 Ω-cm at 300 K) which becomes insulating at $T < 150$ K, to avoid high frequency losses in the substrate. Hence the graphene-substrate capacitance ($C_{chan}$ in Fig 3(a)) can be ignored at low temperatures. After defining contacts with e-beam lithography, we deposit Ti by e-beam evaporation and oxidize it in pure oxygen (~ 0.5 atm) for 4 hours. The oxidized Ti creates a tunnel barrier with a junction resistance of about ~10-100 $k\Omega\cdot\mu m^2$. NbN is then deposited through DC sputtering which provides a large superconducting gap (~2 meV) for confining the hot electrons.

The measurement setup is shown in Fig 3(a). The RF signal which heats the absorber is provided by an Agilent E4422 RF signal generator. A directional coupler couples the input RF signal through the RF/capacitive port of a bias tee into the device under test (DUT). The reflected RF signal from the device, appearing at the capacitive port of the bias tee is amplified by an RF amplifier before being measured by an Agilent E4416 power meter. The DC characteristics of the devices are measured by passing a 10 nA, 13 Hz current (using a Keithley6221 current source) through the DC/inductive port of the bias tee. At the same time we measure the voltage drop using a SR830 lock-in amplifier. At room temperature, a bank of two-stage RC low pass filters is used to heavily filter out the high frequency noise in the DC branch. All of our RF components have their responses limited to frequencies between 2 GHz to 8 GHz, and the measurements are done at 4 GHz where the amplifier has a maximum gain (~26 dB) for the reflected power. All RF components are kept at room temperature in our setup and the connection to the device in the cryostat is through a semi-rigid coaxial cable.

To accurately obtain the RF power absorbed by the device, we calibrated the setup by measuring the gain/loss of each section and by calibrating using 50 Ω, short and open loads, from which the gain/loss of each branch of the circuit is obtained. From the applied and reflected power we can calculate the voltage reflection coefficient ($\Gamma$) for our device. This is given by,

$$|\Gamma|^2 = \frac{P_o}{P_i} \quad (2)$$

where, $P_o$ and $P_i$ represent measured reflected (output) power and applied (input) power in watts respectively at the DUT. From equation (2) and the relation, $P_{abs} = (1 - |\Gamma|^2)P_i$ we can calculate the power absorbed in the device. The voltage reflection coefficient in turn can be used to determine the device impedance as seen by the RF signal:

$$\Gamma = \frac{Z_{device} - Z_{source}}{Z_{device} + Z_{source}}. \quad (3)$$

where $Z_{source} = 50\ \Omega$. The measurement of the absorbed power in the device is reasonably accurate when the device impedance is comparable to 50 Ω.

For an applied power of -61 dBm, the measured temperature dependence of the reflected power is shown in Fig 3(b) (black curve). For $T > T_c$, when the leads are resistive, applied power is mostly reflected back, implying $|\Gamma| \sim 1$. As the leads become superconducting at $T \sim T_c$, the reflected power sharply drops, and after a small dip it settles to a constant value for $T \ll T_c$. This is mirrored in the reflection coefficient magnitude as a sharp drop to a minimum value before slightly turning up and roughly flattening out when $T \ll T_c$. This is in contrast with the DC resistance measurement in which the device resistance increases continuously with decreasing temperature for $T < T_c$ (Fig 3(b) red curve), signaling that the device impedance at high frequencies is very different from the DC resistance.

We consider a lumped parameter model of the device, shown in Fig 3(a) right inset. With C being the contact capacitance of each metal-graphene junction, the device impedance can be written as $Z_{device} = \frac{2R_c Z_c}{R_c + Z_c} + R_L + R_g$ where $R_g$ is the resistance of the graphene channel, $R_L$ is the lead resistance, $R_c$ is the DC contact resistance for each contact and $Z_c = \frac{1}{i\omega C}$. Since titanium oxide has a large dielectric constant (ε~100), a large capacitance between graphene and the superconducting contacts is present. In this case of each 14 $\mu m^2$ size contacts, we estimate this capacitance to be ~12 pF for each contact, which leads to a capacitive impedance of ~3 Ω at the frequency of 4 GHz used in our measurements. This capacitive impedance, which is in parallel with the tunneling/contact resistance of a few kΩ, provides a "short circuit path" for the RF signal. Consequently, the resistance of the graphene channel, which is roughly temperature independent, is the main contribution to RF impedance.

For the RF signal, neglecting the DC contact resistance $R_c$, $Z_{device} \approx 2Z_c + R_L + R_g \equiv 2Z_c + R$, where $R = R_L + R_g$ is the real component of the impedance.

Based on equation (3) with the above mentioned form of $Z_{device}$ we plot $|\Gamma|$ vs. R for a few different values of C (see Fig 3(b) inset). We see that $|\Gamma|$ has a unique minimum value depending on the value of C. By comparing the minimum value of $|\Gamma|$ found based on our experiment with the plot in Fig 3(b) inset, we can find the value of C for our device. We find that the minimum of $|\Gamma|_{min}$~ 0.09 occurs at T~11 K in our experiment. According to our device model, this $|\Gamma|_{min}$~ 0.09 occurs when C~9 pF. This value is reasonably close to the value of ~12 pF estimated above based on the geometry of this device. The slight discrepancy could be due to the fact that the dielectric constant of TiOx may be different at our operation frequency. It can also be seen that this minimum occurs at $R = R_L + R_g = 52\ \Omega$. Experimentally, as the temperature is reduced below the superconducting transition temperature of the contacts, $R_L$ rapidly decreases to zero, and R decreases to a temperature independent value of $R = R_g$. Consequently the reflection co-efficient $|\Gamma|$ increases and then becomes constant. We can solve equation (3) for $R = R_g$ from the value of $|\Gamma|$ at $T < T_c$ and by substituting C~9 pF in $Z_{device} = 2Z_c + R$. This way, we find that $R_g \sim 10\ \Omega$ and see that it is roughly temperature independent as is expected from the extremely weak electron-phonon scattering at low temperatures (Fig 3(c)). This small resistance value may be attributed to the lager aspect ratio $(W/L)$ of this device (~35) and a strong residue doping.

In figure 4(a) we show differential resistance vs. device voltage for a graphene/TiOx/NbN device at different bath temperatures with no RF radiation. We see that with increasing temperature, the differential resistance decreases when the bias voltage is less than the superconducting gap $|V_b| < \Delta_{NbN}$, and is roughly temperature independent when $|V_b| > \Delta_{NbN}$, characteristic of quasiparticle tunneling behavior. Fig 4(b) shows the effect of radiation on this device at a fixed base temperature. These curves are measured by keeping the bath temperature constant at 4.5 K, but with different power levels of RF radiation applied. The range of RF power applied is chosen so that the bath temperature does not show noticeable increase. Similar to what was observed in graphene/TiOx/Al junctions [4], we see from Fig 4(a) and (b) that both sets of curves are identical in shape. By overlapping the dV/dI vs. bias voltage curves shown in Fig 4(a) and (b), we can calibrate the temperature rise in the device due to a given amount of power absorbed.

**Discussion**

The relation between the power absorbed and the electron temperature rise in the TiOx/NbN device is obtained by comparing the dependences of the zero bias resistance on temperature and on radiation power. The resulting cooling power vs. temperature curve is shown in Fig. 4(c). The temperature dependence of the cooling power can be used to understand hot carrier energy relaxation processes in our device. There are several energy relaxation pathways: 1. Heat can dissipate through low frequency photon emission, given by $G_{photon} \approx k_B B$ where $B \ll k_B T_e/h$ is the coupled bandwidth to an impedance-matched load. 2. Hot electrons can diffuse through metallic leads, given by Wiedemann-Franz law, $G_{WF} = \alpha LT/R_s$. Here L is the Lorentz number, $R_s$ is the total DC contact resistance and $\alpha = 4$ is the geometrical factor for the tunneling contacts. The large tunneling resistance, as a result of suppression of quasiparticle tunneling, allows a small diffusion thermal conductance. At $T = 4.5$ K, $R_s = 51$ kΩ gives $G_{WF} \sim 8.6 \times 10^{-12}$ W/K, which is orders of magnitude lower than our measured thermal conductance $G_{measured} \sim 6.7 \times 10^{-10}$ W/K. 3. Since graphene sits on a $SiO_2$ substrate, this can also provide a medium for heat transfer [20]. It would be instructive to build a suspended-graphene device with superconducting tunnel junction (STJ) geometry to study contribution of this channel at low temperatures. For now, we ignore the contribution of substrate and conclude later that this is indeed not a dominant source of cooling in our experiment. 4. Hot electrons can also cool down via scattering with graphene phonons, specifically through acoustic phonons in the temperature range of our experiment [6, 7].

Heat transfer between electron and acoustic phonons at low temperatures follows a power law given by,[9]

$$P = A\Sigma(T_e^{\delta} - T_{ph}^{\delta}) \qquad (4)$$

where A is the area of the graphene flake, Σ is the electron-phonon coupling parameter and $T_e(T_{ph})$ is the electron(phonon) temperature. The magnitude of the coupling constant Σ and the exponent δ have been calculated in ref [8, 9] for a disorder-free graphene flake. In this case the temperature dependence has been predicted and observed to give $\delta = 4$ [1, 2]. This law has been shown to be modified to $\delta = 3$ in the presence of disorder in the graphene flake, although under different mechanisms in different temperature regimes. When $T_{ph} \gg T_{BG}$, ($T_{BG}$ is the Bloch-Gruneisen temperature of graphene given by $k_B T_{BG} = (2s/v_F)E_F$) the presence of disorder affects electron-phonon scattering via the supercollision mechanism [21]. However, when $T_{ph} < T_{BG}$, as is the case in our experiment, a new temperature regime, based on the disorder level present, appears [16]. At these low temperatures, the phonon wavelength becomes comparable to electronic mean free path. This temperature is given by Chen et al. as $T_{dis} = hs/k_B l$, with $l$ being the electron mean free path and $s = 2 \times 10^4$ m/s being the sound velocity in graphene. Below this temperature due to diffusive transport, an electron has longer time to interact with phonons in the graphene channel and according to [16], in the case of deformation potential coupling and in the absence of screening, electron-phonon scattering is enhanced and $\delta = 3$ in equation (4).

As shown in Fig 4(c), experimentally observed temperature dependence of power can be fitted to the relation $P = 0.132A(T_e^3 - T_{ph}^3)$ with an estimated graphene area of $A \sim 100$ $\mu m^2$. Comparing this to equation (4), we extract an electron-phonon coupling of $\Sigma = 132$ $mW/K^3m^2$. Using a typical mean free path value of $l$~20 nm [22] for graphene on a $SiO_2$ substrate, $T_{dis}$ is close to 50 K. Without a back gate, we estimate the Fermi energy in graphene based on resistivity ($R_{square}$) obtained from the RF reflection coefficient measurement discussed previously. The value of $R_{square} \ll 500\ \Omega$ suggests that the graphene channel was sufficiently doped away from the CNP to satisfy the relation $k_F l < 1$, required for the disorder assisted electron-phonon scattering derived in [16]. According to [16], since our observed temperature dependence occurs only for the case of electron-phonon coupling through deformation potential in the weak screening limit, $\Sigma$ is given by,

$$\Sigma = \frac{2\varsigma(3)}{\pi^2} D^2 \frac{E_F}{\hbar^4 \rho_m s^2 v_F^3 l} (k_B T)^3 \qquad (5)$$

where D is the deformation potential, $\rho_M$ is the mass density of graphene and $\varsigma$ denotes the Riemann-zeta function. For a typical graphene flake on $SiO_2$ substrate, a low resistivity of ~350 Ω, which was estimated through RF reflection coefficient, would signify that the Fermi energy is ≥ 100 meV. From this lower bound of $E_F$ and assuming a typical mean free path of 20 nm, we can estimate the deformation potential to be $D \sim 20$ eV. This matches with the theoretical predictions of the deformation potential in graphene to lie between 10-30 eV [23]. A lower bound for this disordered behavior prediction is given by $(s/v_F)T_{dis} \sim 1$ K when mean free path $l \sim 20$ nm. It would be useful for potential applications, including bolometry, to study the electron-phonon interaction below this temperature. Although in graphene devices with non-superconducting contacts, diffusion typically dominates over electron-phonon conduction pathway at lower temperatures; our tunnel-junction scheme would be particularly useful here, allowing us to study low-temperature cooling power in graphene due to phonons. From the relation $NEP_{th} = \sqrt{4k_B T^2 G}$ we calculate the intrinsic thermal noise equivalent power to be $0.8\ \mathrm{fW/Hz^{1/2}}$ at 4.6 K.

As discussed at the beginning, to build a state-of-the-art graphene based bolometer three important issues need to be considered: achieving the electron-phonon cooling bottleneck, impedance matching for high input coupling efficiency and electron temperature readout. In graphene-superconductor Josephson-like weak links (device D1), both the electron-phonon cooling bottleneck and impedance matching have been achieved. However the technique itself requires significant heating to overcome the supercurrent and readout of the electron temperature is not direct. Here we show that for Josephson weak link type of junctions, DC resistance readout for electron temperature thermometry is difficult due to the presence of strong non-linear effect in the sub-gap regime. This, however, does not rule out possibility of detecting the electron temperature through other techniques. In particular, direct measurement of the electron temperature through Johnson noise has been a well-developed technique and has been theoretically modeled to be a promising way for building high performance graphene-superconductor bolometers [3, 18].

The graphene-superconductor junction with a tunnel barrier contact is demonstrated to be a promising scheme for high performance bolometer with a direct resistance readout. The microwave impedance of the devices can be well matched with an antenna to give high coupling efficiency, due to the large tunneling capacitance. The electron temperature can be directly measured through the thermally excited quasiparticle tunneling resistance. The tunnel barrier blocks diffusion of the hot electrons through the contact leads to achieve low thermal conductance and high sensitivity. Although at this time fabrication of a fully oxidized, high quality tunnel barrier on graphene with an ideal exponential R(T) behavior and large capacitance still needs to be developed, we believe that this technical difficulty can be solved, allowing phonon cooling to be the bottleneck for limiting the thermal conductance and the sensitivity of the devices. Further improvement of bolometer sensitivity should reduce the coupling constant $\Sigma$, which can be achieved by lowering the Fermi energy, and using a cleaner graphene flake with longer mean free path, so that electron-phonon coupling is weaker at lower temperatures.

In summary, we have demonstrated that the radiation response of a transparent graphene-superconductor device is dominated by device non-linearity. In contrast, graphene-superconductor tunnel junctions show a predominantly bolometric radiation response. The results shown here demonstrate that graphene STJ bolometers are capable of providing high sensitivity with practical efficiency due to achievable low RF impedance.

The authors thank Dr. Daniel Prober, Chris McKitterick, Dr. K.C Fong and Dr. G. Finkelstein for useful discussions. This work was supported by AFOSR-YIP Award No. FA9550-10-1-0090.

## References

1. Betz, A.C., et al., *Hot Electron Cooling by Acoustic Phonons in Graphene.* Physical Review Letters, 2012. **109**(5): p. 056805.
2. Fong, K.C. and K.C. Schwab, *Ultrasensitive and Wide-Bandwidth Thermal Measurements of Graphene at Low Temperatures.* Physical Review X, 2012. **2**(3): p. 031006.
3. McKitterick, C.B., D.E. Prober, and B.S. Karasik, *Performance of graphene thermal photon detectors.* Journal of Applied Physics, 2013. **113**(4): p. 044512-6.
4. Vora, H., et al., *Bolometric response in graphene based superconducting tunnel junctions.* Applied Physics Letters, 2012. **100**(15): p. 153507.
5. Yan, J., et al., *Dual-gated bilayer graphene hot-electron bolometer.* Nat Nano, 2012. **7**(7): p. 472-478.
6. Bistritzer, R. and A.H. MacDonald, *Electronic Cooling in Graphene.* Physical Review Letters, 2009. **102**(20): p. 206410.
7. Tse, W.-K. and S. Das Sarma, *Energy relaxation of hot Dirac fermions in graphene.* Physical Review B, 2009. **79**(23): p. 235406.
8. Kubakaddi, S.S., *Interaction of massless Dirac electrons with acoustic phonons in graphene at low temperatures.* Physical Review B, 2009. **79**(7): p. 075417.
9. Viljas, J.K. and T.T. Heikkilä, *Electron-phonon heat transfer in monolayer and bilayer graphene.* Physical Review B, 2010. **81**(24): p. 245404.
10. Du, X., et al., *Graphene-based Bolometers.* arXiv preprint arXiv:1308.4065, 2013.
11. Karasik, B.S., C.B. McKitterick, and D.E. Prober, *Prospective performance of graphene HEB for ultrasensitive detection of sub-mm radiation.* J. Low Temp Phys., submitted, 2013.
12. Efetov, D.K. and P. Kim, *Controlling Electron-Phonon Interactions in Graphene at Ultrahigh Carrier Densities.* Physical Review Letters, 2010. **105**(25): p. 256805.
13. Borzenets, I.V., et al., *Phonon Bottleneck in Graphene-Based Josephson Junctions at Millikelvin Temperatures.* Physical Review Letters, 2013. **111**(2): p. 027001.
14. Breusing, M., C. Ropers, and T. Elsaesser, *Ultrafast Carrier Dynamics in Graphite.* Physical Review Letters, 2009. **102**(8): p. 086809.
15. Tinkham, M., *Introduction to superconductivity*. Second ed. 2012: Dover Publications.
16. Chen, W. and A.A. Clerk, *Electron-phonon mediated heat flow in disordered graphene.* Physical Review B, 2012. **86**(12): p. 125443.
17. McKitterick, C.B., et al., *Graphene microbolometers with superconducting contacts for terahertz photon detection.* arXiv preprint arXiv:1307.5012, 2013.
18. Fong, K.C., et al., *Measurement of the Electronic Thermal Conductance Channels and Heat Capacity of Graphene at Low Temperature.* Physical Review X, 2013. **3**(4): p. 041008.
19. Andreev, A.F., Sov.Phys.JETP, 1964. **19**(1228).
20. Freitag, M., T. Low, and P. Avouris, *Increased Responsivity of Suspended Graphene Photodetectors.* Nano Letters, 2013. **13**(4): p. 1644-1648.
21. Song, J.C.W., M.Y. Reizer, and L.S. Levitov, *Disorder-Assisted Electron-Phonon Scattering and Cooling Pathways in Graphene.* Physical Review Letters, 2012. **109**(10): p. 106602.
22. Du, X., I. Skachko, and E.Y. Andrei, *Josephson current and multiple Andreev reflections in graphene SNS junctions.* Physical Review B, 2008. **77**(18): p. 184507.
23. Hwang, E.H. and S. Das Sarma, *Acoustic phonon scattering limited carrier mobility in two-dimensional extrinsic graphene.* Physical Review B, 2008. **77**(11): p. 115449.

**List of Tables**

**Table 1** devices discussed in this article: Note: Similar to Pd/Al contacts, non-linearity dominated response was observed in transparent graphene-superconductor junctions with Ti/Nb and Ti/NbN contacts as well which is not shown here.

**Figure Captions**

**Fig 1** Cartoon schematics for the two types of superconductor-graphene-superconductor (SGS) devices with a transparent S-G interface (bottom) and a tunnel barrier S-G interface (top).

**Fig 2** Non-linear response in graphene/Pd/Al junctions (a) Differential resistance vs. junction voltage curves taken at different temperatures for $V_g - V_{CNP} = 15$ V with no radiation applied. *Inset:* differential resistance vs. junction voltage curves with different back-gate voltages, with no radiation applied. (b) Radiation induced non-linear response. Upper panel shows simulation results, lower panel shows measured response. $I_{RF} = 0,50,120,300,450$ nA was used for simulation, in the increasing radiation direction.

**Fig 3** Impedance matching: (a) Measurement setup schematic All the RF components are kept at room temperature and the connection to the device at 4 K is through a semi-rigid coaxial cable. Upper-right corner: lumped parameter model of superconducting tunnel contacts on graphene. Lower-right corner: SEM image of a typical graphene-NbN device. (b) Left axis: Reflected power variation with temperature (with -48.9 dBm amplifier input noise subtracted) Right axis: zero bias differential resistance variation with temperature (both curves at an applied signal power of -61 dBm). *Inset:* Calculated reflection coefficient variation with resistance at a few different contact capacitance values. (c) Left Axis: Measured DC resistance at different temperatures at an applied signal power of -61 dBm Right axis: Calculated high frequency device resistance.

**Fig. 4** Bolometric detection in NbN tunnel devices: (a) Differential resistance vs. junction voltage curves taken at different temperatures with no radiation applied. (b) Differential resistance vs. junction voltage curves with different amounts of RF power applied at 4.5 K bath temperature. (c) Calibration of induced temperature change by absorbed power in graphene/TiOx/NbN device, fitted to a $T^3$ temperature dependence indicating a disorder-induced modification of electron-phonon cooling channel.

| Devices | D1 | D2 |
|---|---|---|
| **S-G Interface** | Transparent | Tunnel barrier |
| **Contact Metals** | Pd/Al | TiOx/NbN |
| **Radiation Response** | Non-linear | Bolometric |
| **Blocking of hot carrier diffusion** | Andreev Reflection | Quasiparticle tunneling suppression |
| **Temperature range studied** | 230 mK-1 K | 4-8 K |
| **Substrate used** | Conducting Si + $SiO_2$(300 nm) | High resistivity Si + $SiO_2$(500 nm) |
| **Radiation coupling** | 1.3 GHz radiation through free space | 4 GHz radiation through coaxial cable |
| **Contact area** | 6 $\mu m^2$ | 14 $\mu m^2$ |

Table 1

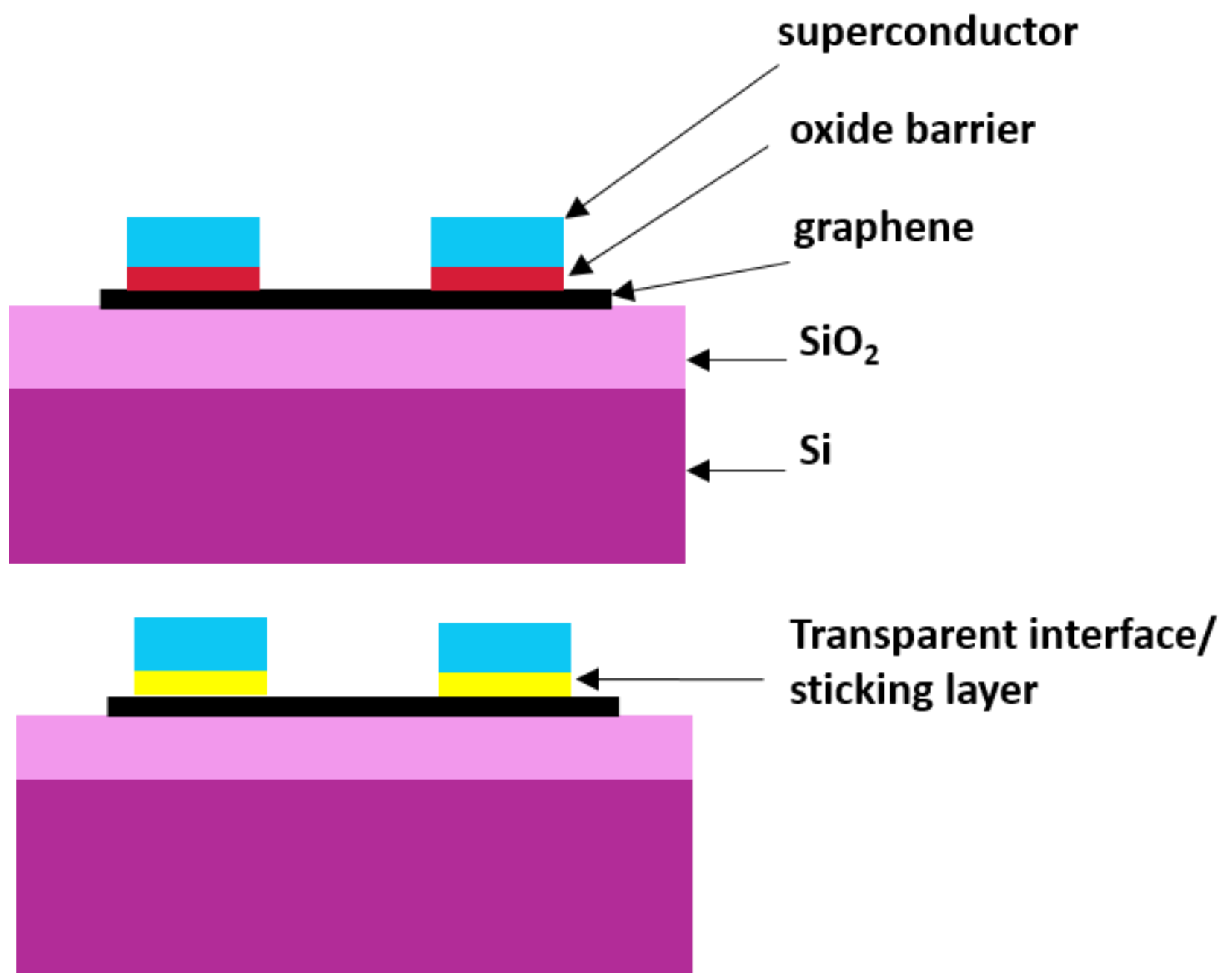

Figure 1

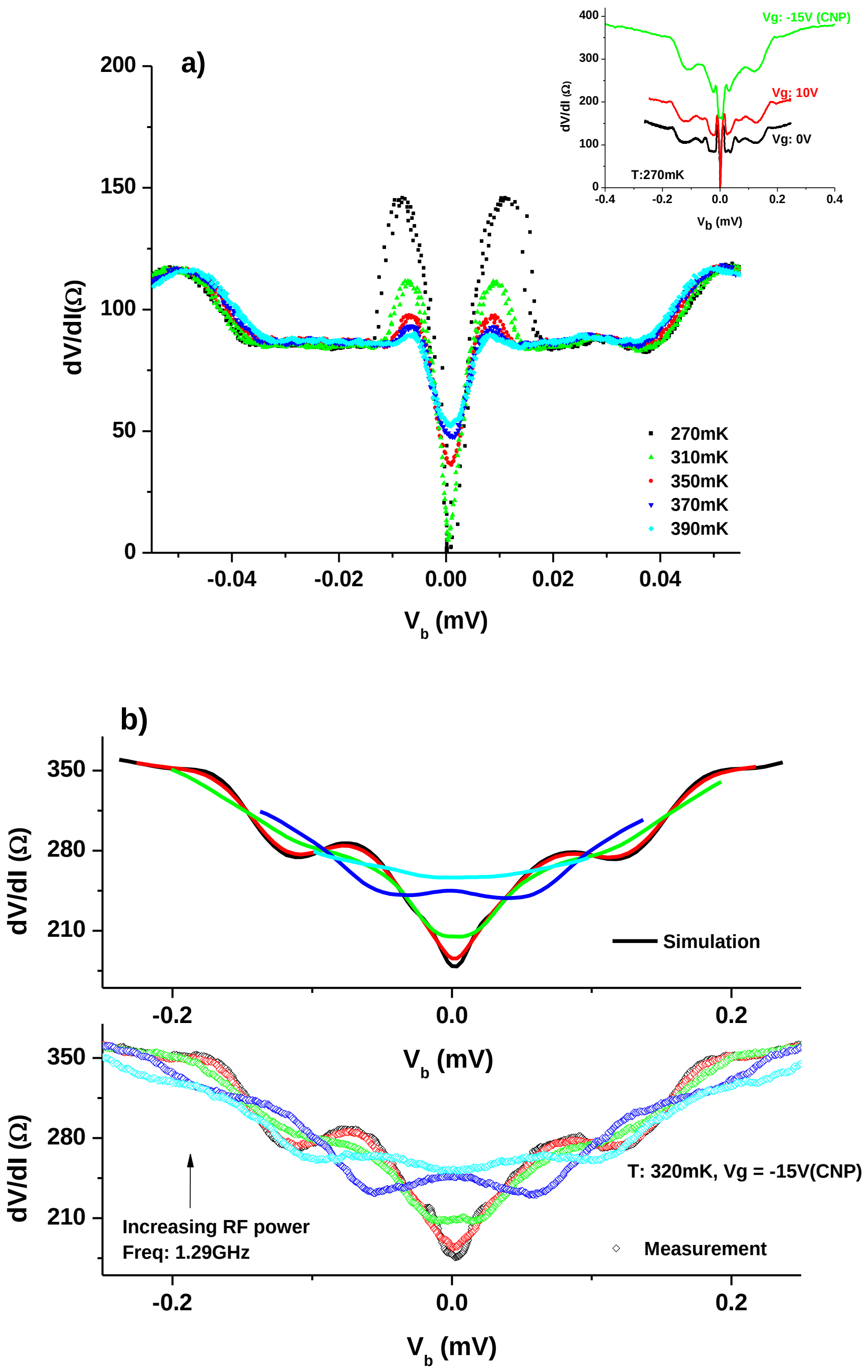

Figure 2

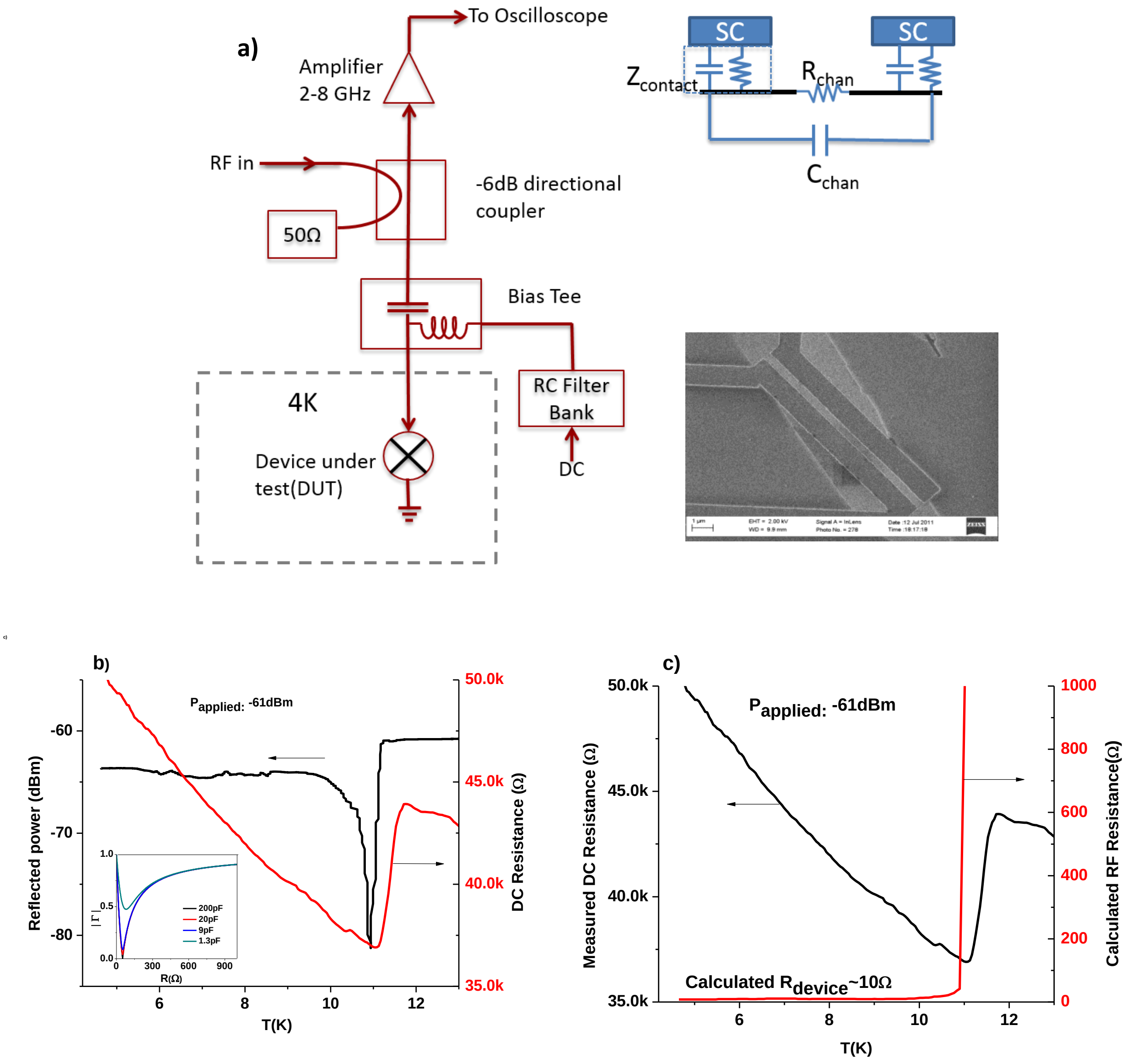
a)
To Oscilloscope
Amplifier
2-8 GHz
RF in
50Ω
-6dB directional
coupler
Bias Tee
RC Filter
Bank
DC
4K
Device under
test(DUT)
SC
SC
Zcontact
Rchan
Cchan
b)
Papplied: -61dBm
Reflected power (dBm)
DC Resistance (Ω)
T(K)
-60
-70
-80
50.0k
45.0k
40.0k
35.0k
6
8
10
12
1.0
0.5
0.0
|Γ|
0
300
600
900
R(Ω)
200pF
20pF
9pF
1.3pF
c)
Papplied: -61dBm
Measured DC Resistance (Ω)
Calculated RF Resistance(Ω)
Calculated Rdevice~10Ω
T(K)
50.0k
45.0k
40.0k
35.0k
1000
800
600
400
200
0
6
8
10
12

Figure 3

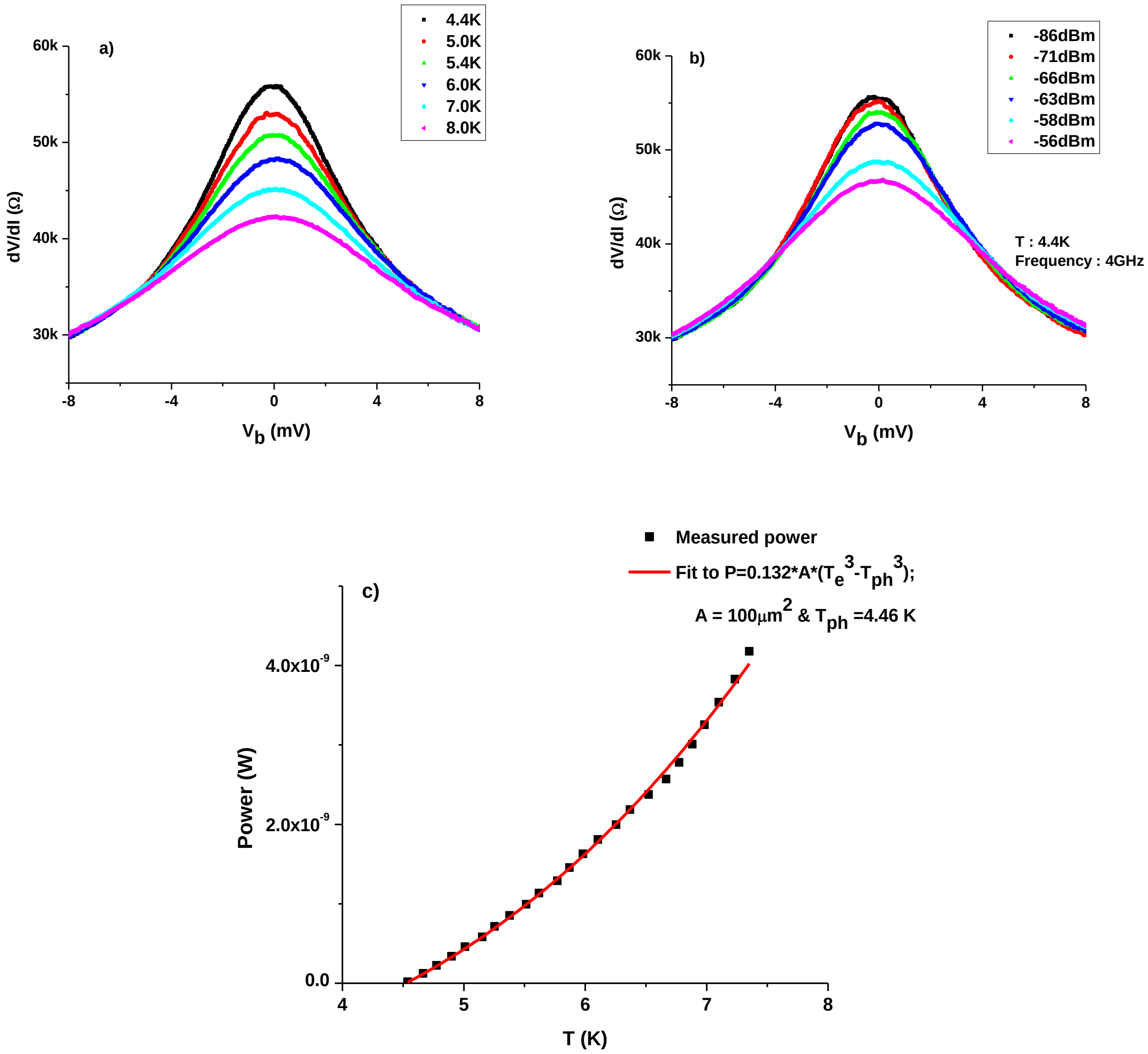

Figure 4